\documentclass{aa}  

\usepackage{graphicx}
\usepackage{txfonts}
\usepackage{amsmath,color}
\begin{document}

   \title{H{\Large I} properties and star formation history of\\ a fly-by pair of blue compact dwarf galaxies}
   \titlerunning{HI properties and SF history of a fly-by pair of BCDs}
   \author{Jinhyub Kim\inst{1},
           Aeree Chung\inst{1,}\inst{2,}\inst{3}\thanks{Email: {\tt achung@yonsei.ac.kr}},
           O. Ivy Wong\inst{4},
           Bumhyun Lee\inst{1},
           Eon-Chang Sung\inst{5},
           Lister Staveley-Smith\inst{4,}\inst{6}
          }
   \authorrunning{Kim et al.}
   \institute{Department of Astronomy, Yonsei University, 50 Yonsei-ro, Seodaemun-gu, Seoul 03722, Korea \and
             Yonsei University Observatory, Yonsei University, 50 Yonsei-ro, Seodaemun-gu, Seoul 03722, Korea \and 
             Joint ALMA Observatory, Alonso de C$\acute{\rm o}$rdova 3107, Vitacura, Santiago, Chile \and             
             International Centre for Radio Astronomy Research (ICRAR), University of Western Australia, 35 Stirling Highway, Crawley, Western Australia 6009, Australia \and
             Korea Astronomy and Space Science Institute 776, Daedeokdae-ro, Yuseong-gu, Daejeon 34055, Korea \and 
             ARC Centre of Excellence for All-sky Astrophysics (CAASTRO) 
             }
   \date{Received oooooo; accepted oooooo}

% \abstract{}{}{}{}{} 
% 5 {} token are mandatory

  \abstract
{A fly-by interaction has been suggested to be one of the major explanations for enhanced star formation in blue compact dwarf (BCD) galaxies, yet no direct evidence for this scenario has been found to date. In the H{\sc i} Parkes all-sky survey (HIPASS), ESO~435$-$IG~020 and ESO~435$-$~G~016, a BCD pair were found in a common, extended gas envelope of atomic hydrogen, providing an ideal case to test the hypothesis that the starburst in BCDs can be indeed triggered by a fly-by interaction. Using high-resolution data from the Australia Telescope Compact Array (ATCA), we investigated H{\sc i} properties and the spectral energy distribution (SED) of the BCD pair to study their interaction and star formation histories. The high-resolution H{\sc i} data of both BCDs reveal a number of peculiarities, which are suggestive of tidal perturbation. Meanwhile, 40\% of the HIPASS flux is not accounted for in the ATCA observations with no H{\sc i} gas bridge found between the two BCDs. Intriguingly, in the residual of the HIPASS and the ATCA data, ~10\% of the missing flux appears to be located between the two BCDs. While the SED-based age of the most dominant young stellar population is old enough to have originated from the interaction with any neighbors (including the other of the two BCDs), the most recent star formation activity traced by strong H$\alpha$ emission in ESO~435$-$IG~020 and the shear motion of gas in ESO~435$-$~G~016, suggest a more recent or current tidal interaction. Based on these and the residual emission between the HIPASS and the ATCA data, we propose an interaction between the two BCDs as the origin of their recently enhanced star formation activity. The shear motion on the gas disk, potentially with re-accretion of the stripped gas, could be responsible for the active star formation in this BCD pair.}

   \keywords{galaxies-- individual: ESO~435$-$IG~020, ESO~435$-$~G~016 -- galaxies: interactions -- galaxies: ISM -- galaxies: kinematics and dynamics -- galaxies: starbursts}

   \maketitle
%
%-------------------------------------------------------------------

\section{Introduction}

Blue compact dwarf galaxies (BCDs) are low-mass systems characterized by active star formation in compact regions and strong narrow emission lines superposed on a flat stellar continuum \citep[e.g.,][]{salzer89,kehrig04}. They were originally considered as either nascent or immature galaxies that had recently undergone intense bursts of star formation \citep{searle73}. However, it turns out that they also contain stellar populations older than $\sim$1~Gyr as shown by a number of optical and near-IR studies \citep{thuan83,papa96,aloisi99}, implying that BCDs are little galaxy fossils where starbursts have been recently triggered.

The low metallicity of BCDs \citep[$1/50 \la Z/Z_\odot \la 1/10$;][]{ko00} also suggests that they have not always maintained their current star formation rate \citep[$\sim$0.1~$M_\odot$ yr$^{-1}$;][]{hopkins02}. Indeed, the neutral hydrogen gas mass found in very gas-rich BCDs is typically $10^8M_\odot$ \citep{tm81}, which will require 1 Gyr to be consumed entirely if they keep up with the current star formation rate \citep{thuan83}. Instead, a number of models have shown that dwarf galaxies may go through a {``BCD phase''} due to the gas infall \citep[e.g.,][]{verbeke14}, merging with another dwarf galaxy \citep[e.g.,][]{dimatteo07,bekki08,cloet14}, or fly-by interaction \citep[e.g.,][]{icke85,pustilnik01}.

Those scenarios have been also tested observationally in a wide range of wavelengths from optical through to radio. For example, merging signatures or the presence of companions have been confirmed in H$\alpha$ \citep[e.g.,][]{ostlin99,me00,le09} and deep optical imaging studies \citep[e.g.,][]{rich12}. In addition, H{\sc i} imaging has been particularly useful in finding the evidence for tidal interactions in BCDs \citep[e.g.,][]{vanZee98,chengalur06,ekta06,lelli12,lopez12,nidever13}. {Approximately 57\% of BCDs have been found with gas-rich but optically faint dwarf nearby companions \citep[][]{taylor97}, and the fraction increases up to 80\% when BCDs with signatures of tidal interaction including merging are included \citep{pustilnik01}.} All these observations support the hypothesis that the bursts of star formation in BCDs can be induced externally.

On the other hand, there are cases of BCD pairs with separation larger than a few tens~kpc such as UM 461/462 \citep[$d\sim$68~kpc;][]{taylor95}, ESO 338-004/004B \citep[$d\sim$72~kpc;][]{cannon04}, ESO 400-043/043B \citep[$d\sim$70~kpc;][]{ostlin99, ostlin01}, and IRAS 08339+6517/2MASX J08380769+6508579 \citep[$d\sim$56~kpc;][]{cannon04}. For those BCD pairs, it is more difficult to prove that the two galaxies are or have been tidally interacting with each other. And even if this is so, how their "fly-by" interaction could lead to the recent active star formation is still questionable. Considering that there is quite a high fraction of BCD galaxies that are not in the process of merging \citep[$\sim$60\%,][]{pustilnik01,sung02}, it is important to show that they can indeed originate from fly-by interactions.

\begin{figure}
 \begin{center}
 \includegraphics[bb=70 170 525 690,width=0.5\textwidth]{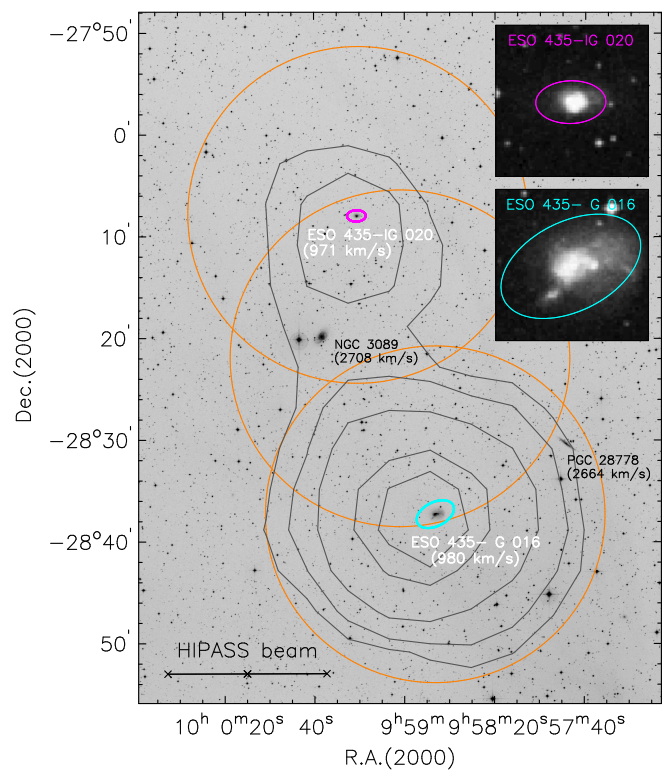}
 \end{center}
 \caption{HIPASS H{\sc i} column density contours are overlaid on the DSS2 (Digital Sky Survey 2) red optical image. Two small colored ellipses represent the optical size, position angle (PA), and the inclination of our targets, taken from \citet[RC3]{RC3}. ESO catalog names of our BCDs are presented in white with their systemic velocity. Two other optically identified galaxies in this field, which are background systems, are also annotated in black. The HIPASS H{\sc i} contour levels are (3, 6, 12, 24, 30) $\times$ 10$^{18}$~cm$^{-2}$. The primary beam of the Parkes telescope (14\arcmin) is shown at the lower left corner. Three ATCA pointings, each of which corresponds to the size of the ATCA primary beam ($\approx 33.1$ arcmin in diameter) are indicated by the orange line. On the top right corner, the zoom-in DSS2 blue images of individual galaxies are presented. Each panel is the size of $2\times2~$arcmin$^2$. Again, the colored ellipse represents the optical size and PA of each galaxy \citep[]{RC3}.}
 \label{fig_hipass}
\end{figure}

A pair of BCDs, ESO~435$-$IG~020 and ESO~435$-$~G~016, separated by $\sim$80~kpc in projection at similar distances \citep[$\sim$9~Mpc;][]{11HUGS_Ha}, is one such example. Based on the fact that both galaxies appear to be highly disturbed \citep{me00,rj04} in the optical observations, \citet{sung02} have classified this pair as {``detached interacting BCDs''}, proposing that their peculiar optical morphology originates from tidal interaction with each other. Intriguingly, a large neutral hydrogen gas envelope has been revealed in the H{\sc i} Parkes all sky survey \citep[HIPASS;][]{HIPASS1}, which is found to be covering both galaxies (see Figure \ref{fig_hipass}). Their optical morphologies and the common H{\sc i} envelope make this pair a good candidate for the case of a BCD pair with fly-by interaction origin.

Considering the distance between the two galaxies, however, it is difficult to conclude whether H{\sc i} gas associated with individual galaxies simply appears to be connected due to the large HIPASS beam, which is comparable to the separation of the pair on the sky, or whether they are indeed an interacting pair. In order to test the hypothesis that 1) these two dwarfs are going (or have gone) through a fly-by interaction, and 2) their interaction is the cause of their recent star formation activity, we followed up this region in H{\sc i} with a higher resolution using the Australia Telescope Compact Array (ATCA). In this work, we probe detailed H{\sc i} morphology and kinematics of the ESO~435$-$IG~020/016 pair and their star formation properties to investigate the role of a fly-by interaction as one of the important origins for quite a large fraction of the BCD population including our targets. 

This paper is organized as follows. In Section 2, general properties of ESO~435$-$~G~016/020 pair are introduced. In Section 3, the ATCA H{\sc i} observation and data reduction are described. In Section 4, the H{\sc i} morphology and kinematics of individual galaxies are presented.  In Section 5, the presence of intergalactic H{\sc i} between the pair is inspected. Also, the star formation and interaction histories of the pair are discussed. In Section 6, we summarize and conclude. 

%--------------------------------------------------------------------

\begin{table*}
\begin{center}
 \caption{Optical properties of the pair.}
 \label{tbl_genpro}
 \begin{tabular}{@{}lccccccccc}
  \hline\hline
 \noalign{\vspace{1mm}}
  Galaxy & R.A. & DEC.
        & Distance & cz & $M_{B}$ & $M_{*}$ & $D_{25}$ & $r_{25}$ & P.A. \\
        & (J2000) & (J2000) & Mpc & km~s$^{-1}$ & mag & $10^9~M_\odot$ & {arcsec} & ($b$/$a$)$_{25}$ & {deg} \\
 \noalign{\vspace{1mm}}
  \hline
 \noalign{\vspace{1mm}}
  ESO~435$-$IG~020 & $09^h59^m21.2^s$ & $-28{\degr}08'00''$ & 9.0 & 971 & $-15.67$ & 0.26 & 49.90 & 0.71 & 91.7 \\
  ESO~435$-$~G~016 & $09^h58^m46.2^s$ & $-28{\degr}37'19''$ & 9.1 & 980 & $-16.46$ & 2.34 & 99.60 & 0.68 & 117 \\
 \noalign{\vspace{1mm}}
  \hline
 \end{tabular}
\end{center}
\footnotesize{{\bf Notes.} Information on the galaxy coordinates (R.A., DEC.) and radial velocities comes from \citet[][RC3]{RC3}. $B-$band luminosities, stellar masses, optical extents, and distances are taken from \citet{11HUGS_Ha}, and \citet{fd11}. Distances are corrected for the relative motion to the centroid of the Local Group \citep{11HUGS_Ha}.}
\end{table*}

\section{ESO~435$-$IG~020 and ESO~435$-$~G~016}
Both galaxies are small (a few kpc in diameter) and faint ($M_B\lesssim-18.5$~mag) with strong, sharp, and narrow emission lines on a blue continuum, and both are classified as blue compact dwarf (BCD) galaxies \citep{sung02}. They are found at a similar redshift ($\sim970-980~$km~s$^{-1}$) with an angular separation of $\sim30~$arcmin, which corresponds to $\sim$80~kpc, adopting a 9~Mpc distance to the pair \citep{11HUGS_Ha}.

ESO~435$-$IG~020 has a compact core surrounded by an asymmetric, elongated stellar envelope. It was classified as an ``interacting'' system in the ESO atlas, as hinted by the catalog nomenclature \citep[i.e., IG=interacting galaxy or galaxies][]{ESOatlas}. \citet{doublier99} also found its brightness distribution to follow $r^{1/4}$ rather than an exponential profile, which is suggestive of (partial) violent relaxation after merging.

ESO~435$-$~G~016 also has a diffuse stellar envelope which is off-center from the bright component. This galaxy has two sharp stellar tails; one inside the diffuse component in the west and a more distinctive tail pointing southeast. Based on these features, \citet{rj04} identified this system as a merger remnant (it is referred to as AM~0956$-$282 in their work), although it was not classified as ``interacting'' in the ESO catalog \citep{ESOatlas}. However, this galaxy is distinct from the rest of their sample in size and luminosity. It is the smallest among the sample of \citet{rj04}, being less luminous by 1.75 mag compared to the next faint galaxy in their sample, and fainter by 3 mag or more than the rest of the sample. Therefore, even if ESO~435$-$~G~016 is a merger remnant, it is likely to be the result of merging between tiny galaxies like ESO~435$-$IG~020, which is even smaller.

Alternatively, as \citet{sung02} suggested, the peculiarities of ESO~435$-$IG~020 and ESO~435$-$~G~016 could be the result of tidal perturbation with each other. \citet{sung02} also proposed that fly-by interaction between the two galaxies could be the origin of their recent bursts of star formation; we aim to investigate  this possibility in this work. General properties of the pair can be found in Table~\ref{tbl_genpro}.

\section{ATCA observation and data reduction}
The H{\sc i} observations were conducted on March 27 and 28 2002 with the ATCA in EW367 array configuration (project ID of C1024). The ATCA primary beam is 33.1 arcmin at 1416 MHz, which is comparable to the angular distance between the pair of galaxies of $\sim$30.3 arcmin. With a single pointing, the sensitivity at the locations of the galaxies hence drops by a factor of $\sim$1.8, and therefore we had two extra pointings centered on each galaxy in order to keep the desired sensitivity toward the pair. The observing frequency was centered on 1416 MHz with a total bandwidth of 8 MHz, which is configured into 512 channels, yielding a channel width of 3.3 km~s$^{-1}$.

We calibrated the data with {a software package, {\em Miriad}} \citep{sault95}. For flux and bandpass calibration, PKS 1934-638 (14.87 Jy at the observing frequency) was used. As a phase calibrator, we observed PKS 1015-314 every 45 min during our observations. Before making an H{\sc i} cube, the continuum was subtracted by applying a linear fit to the $uv$ data for a range of line free channels at both ends of the band. 
%After the calibration and the continuum subtraction, dirty maps of individual points were generated, which then were combined into one and deconvolved altogether following the prescription suggested by \citet{cornwell88} for extended emission. 
A final H{\sc i} cube was generated using an {\em AIPS}\footnote{{Astronomical Image Processing System, {\em http://www.aips.nrao.edu}}} command {\tt IMAGR}. {In order to maximize the sensitivity while keeping a reasonable synthesized beam size and shape, we set {\rm robust=0} in {\tt IMAGR}, the middle of natural and uniform weighting scheme \citep{briggs95}.} Three points were stitched together by {\tt FLATN} and corrected for the sensitivity loss corresponding to the primary beam shape. The final cube was smoothed to 5~km~s$^{-1}$ of velocity resolution to achieve a higher signal-to-noise ratio. 

\begin{figure}
 \begin{center}
 \includegraphics[bb=40 260 565 590,width=0.5\textwidth]{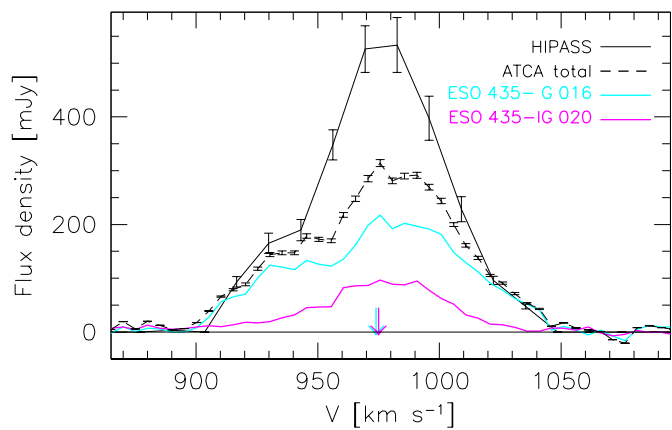}
 \end{center}
 \caption{Comparison of H{\sc i} spectra. Flux densities of individual galaxies are shown in color and the integrated density of the two galaxies is indicated by the dashed black line. A black solid line represents the HIPASS flux density of the pair. The systemic velocity of each galaxy determined within a stellar disk is shown by a downward arrow of the same color as the H{\sc i} spectrum of galaxies. }
 \label{fig_hiprof}
\end{figure}

\begin{table*}
\begin{center}
 \caption{Measured H{\sc i} properties of the targets.}
 \label{tbl_hiprop}
 \begin{tabular}{@{}lccccccccccc}
  \hline\hline
 \noalign{\vspace{1mm}}
  Galaxy & $\Delta_{\rm RA}$ & $\Delta_{\rm DEC}$ & $S_{\rm HI}$ & $M_{\rm HI}$ & $W20$ & $W50$ & $V_{\rm HI}$ & $M_{\rm HI}/L_B$ \\
         &\multicolumn{2}{c}{-- arcsec --} & Jy~km~s$^{-1}$ & $10^8~M_\odot$ & \multicolumn{3}{c}{----- km~s$^{-1}$ -----} & $M_\odot/L_\odot$ \\
 \noalign{\vspace{1mm}}
  \hline
 \noalign{\vspace{1mm}}
ESO 435$-$IG 020 & 7 & 16 &  5.71$\pm$0.19 & 1.11 & 100.1 & 63.4 & 973.0 (974.0) & 0.38 \\
ESO 435$-$~G 016 & 5 &  8 & 15.97$\pm$0.10 & 3.12 & 133.9 & 93.7 & 973.2 (974.9) & 0.52 \\
 \noalign{\vspace{1mm}}
  \hline
 \end{tabular}
\end{center}
\footnotesize{{\bf Notes.} The 2nd and 3rd columns ($\Delta_{\rm RA}$ and $\Delta_{\rm Dec}$) indicate the offset between the optical center and the peak of the H{\sc i} emission. For H{\sc i} mass, the distances in Table~\ref{tbl_genpro} have been applied. The H{\sc i} velocity defined within the symmetric part of the gas disk is given within the parenthesis.}
\end{table*}

\begin{figure*}
 \begin{center}
 \includegraphics[bb=60 190 605 735,width=1\textwidth]{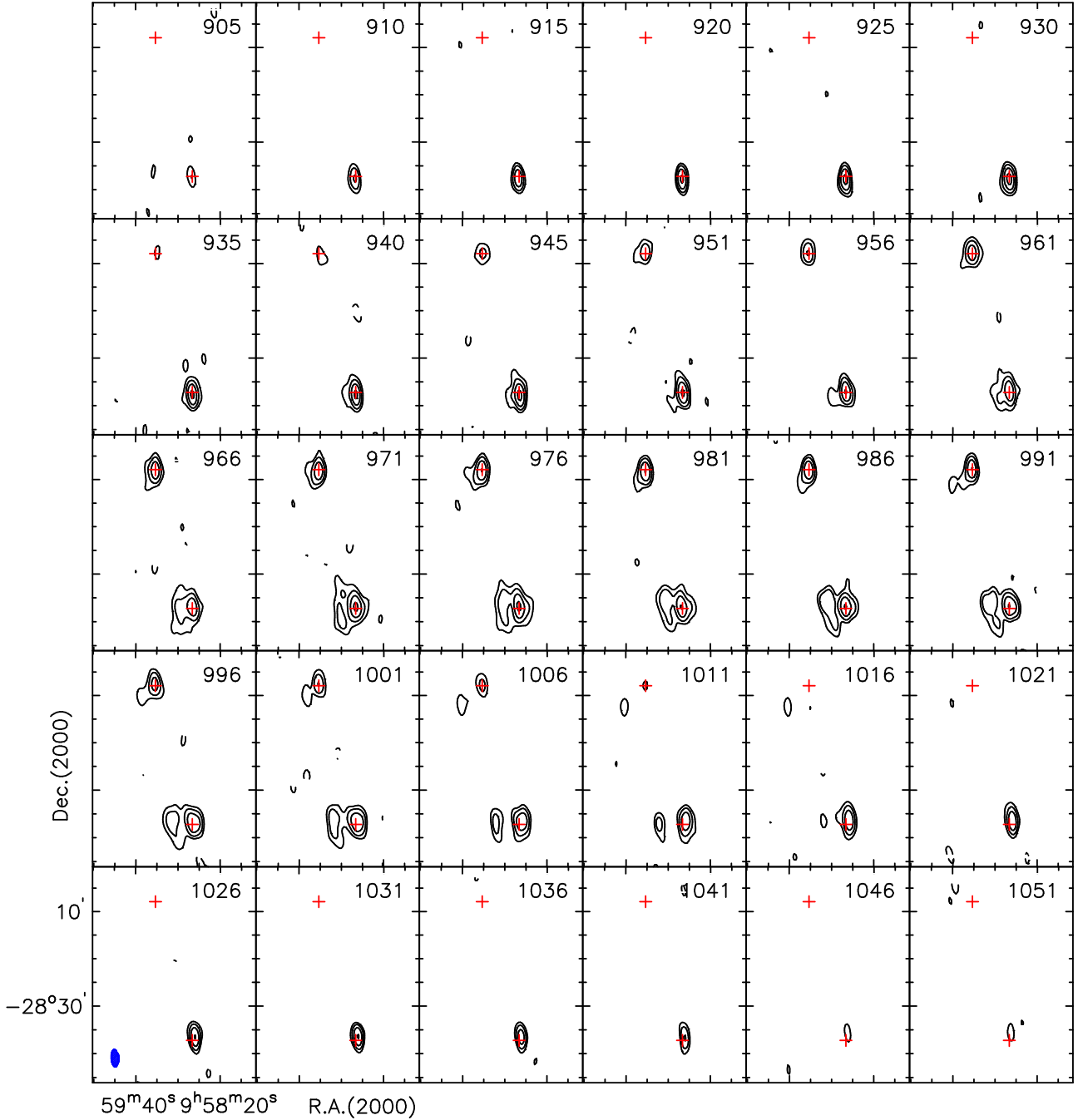}
 \end{center}
 \caption{ATCA H{\sc i} channel maps. H{\sc i} contour levels are ($-6$, $-3$, 3, 6, 12, 24, ...) $\times$ $\sigma$ (negative values in dashed line), where $\sigma=$1.2~mJy~beam$^{-1}$ per 5~km~s$^{-1}$ channel. Each panel is centered on the middle of two galaxies with the size of $\sim60\times120$~kpc$^2$ assuming a distance of 9 Mpc. Red crosses represent the optical center of individual galaxies. The velocity of the channel is shown on the top right of each panel in km~s$^{-1}$. The ATCA synthesized beam (218 arcsec $\times$ 85 arcsec) is indicated by a blue solid ellipse on the bottom left channel with its coordinates.}
 \label{fig_atcach}
\end{figure*}

\begin{figure*}
 \begin{center}
 \includegraphics[bb=100 210 530 585,width=1\textwidth]{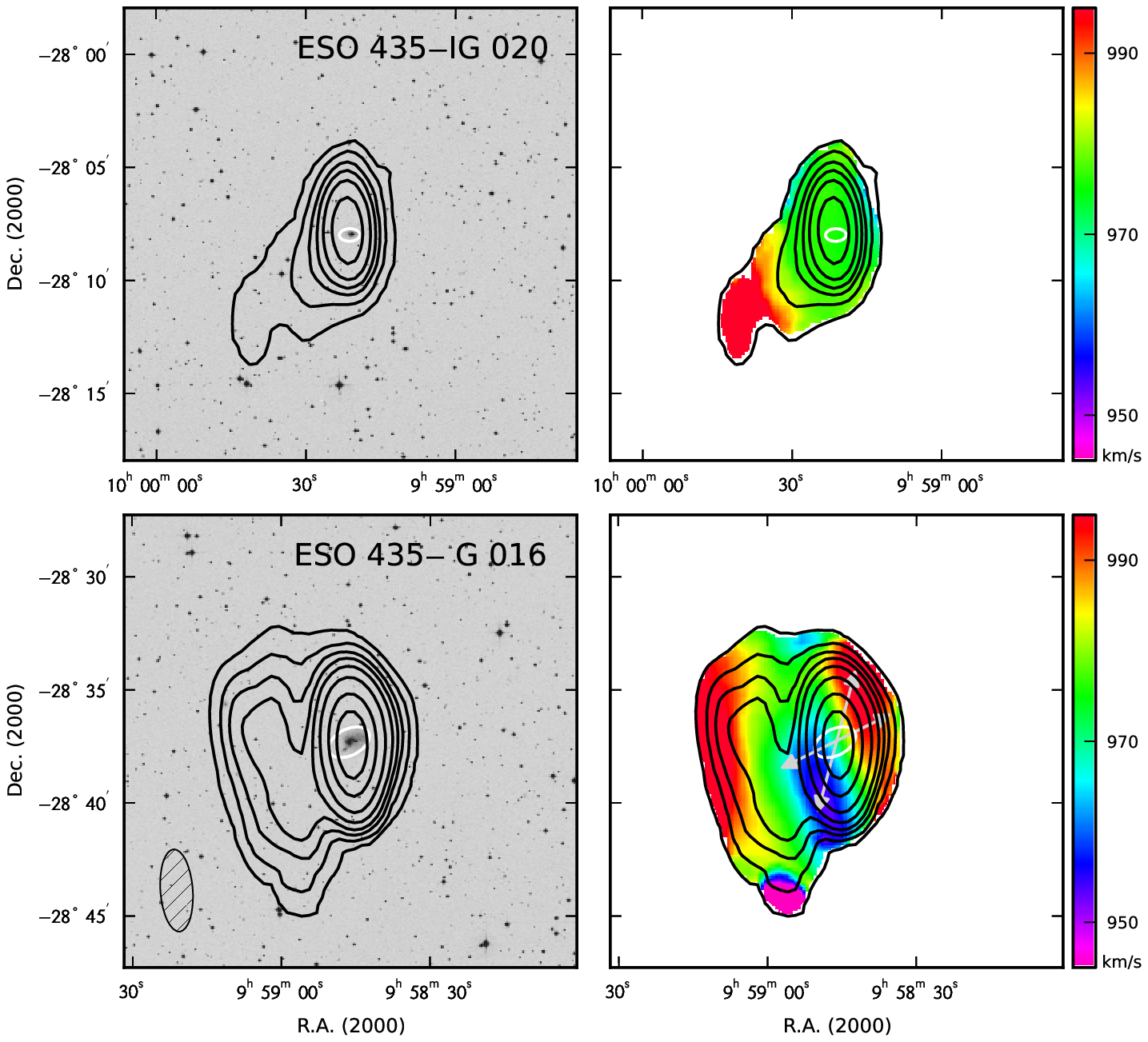}
 \end{center}
 \caption{H{\sc i} intensity map overlaid on the DSS blue image (left) and velocity field (right). The H{\sc i} contour levels are (0.6, 1.8, 3, 4.2, 6, 9.6, 16.8, 25.8) $\times~10^{19}~$cm$^{-2}$. The synthesized beam of our ATCA observation is shown on the bottom left corner and the velocity scale is shown by a color bar on the top right panel. We note that the velocity fields were generated by taking the first moments of the intensity-weighted channels. As the result, the maximum velocity of the northwest end in ESO 435- G 016 appears to be lower (~1000 km s$^{-1}$) than the highest velocity of the channel with emission ($\sim$1050 km~s$^{-1}$). Two gray lines are presented to indicate the optically defined position angle (which agrees with the position angle of the inner H{\sc i} disk) and the position angle of the outer H{\sc i} disk. Further details are discussed in Section~5.3.}
%The dotted lines show the position angle of the xv slices presented in Figure~\ref{fig_atcaxv}. }
 \label{fig_atcamo}
\end{figure*}

The total intensity H{\sc i} map and the intensity weighted velocity field were produced using the task {\tt MOMNT} in {\em AIPS}. It allows us to create a smoothed mask which helps to maximize the signal-to-noise ratio in the integrated map. For a mask, we adopted Gaussian and Hanning smoothing schemes in the spatial and velocity domains, respectively. 
%To probe kinematics, we also created position-velocity diagrams (PVDs) using {\tt KPVSLICE} implemented in {\em KARMA}. The slices were made at the optically defined center along the optical major axes (position angle of 117{\degr} for ESO~435$-$~G~016 and 91.7{\degr} for ESO~435$-$IG~020). In addition, another cut along the position angle of 138{\degr} was made to investigate the motion of the gas tail associated with ESO~435$-$IG~020. The widths of slices are basically $\sim$10\arcsec corresponding to one spatial pixel.

\section{Results}
\subsection{Integrated H{\small I} flux}
In the ATCA data, the pair is resolved into separate structures with the total H{\sc i} flux of ($15.9\pm0.1$) Jy~km~s$^{-1}$ and ($5.7\pm0.2$) Jy~km~s$^{-1}$ associated with ESO~435$-$~G~016 and ESO~435$-$IG~020, respectively. For ESO~435$-$~G~016, the ATCA flux is only 58\% of the previous single-dish measurement \citep[$27.6\pm3.1$ Jy~km~s$^{-1}$, Parkes telescope;][]{koribal04}, while for ESO~435$-$IG~020, we recover all the single-dish flux \citep[$5.4\pm0.5$~Jy~km~s$^{-1}$, Nan\c{c}ay telescope;][]{theu05}. The H{\sc i} profiles from our ATCA observations are compared with the HIPASS H{\sc i} spectrum in Figure~\ref{fig_hiprof}. The total ATCA flux of the pair is $\sim61$\% of the HIPASS flux ($\approx 35$~Jy~km~s$^{-1}$). 

Indeed, the H{\sc i} column density sensitivity of our ATCA observation is not quite comparable to the HIPASS sensitivity (HIPASS versus ATCA $\approx3\times10^{18}~$cm$^{-2}$ vs. $1\times10^{19}~$cm$^{-2}$ in 3~$\sigma$ for $\Delta=50~$km~s$^{-1}$), and hence most faint emission is likely to be missed from the ATCA data. In addition, the largest angular structure detectable using the ATCA EW367 at this wavelength is $\sim16$ arcmin or so, while the H{\sc i} envelope covering the pair found in the HIPASS image is much larger in extent ($\sim50$ arcmin), and the very extended features could have been resolved out. The presence of intergalactic gas is tested and further discussed in Section 5.

%\begin{figure}
% \begin{center}
% \includegraphics[bb=45 195 310 512,width=0.45\textwidth]{fig_atcaxv.ps}
% \end{center}
% \caption{Top) The position-velocity slice of ESO~435$-$IG~020. The cut was made through the optical center and the tip of the south-east tail (PA$=133^\circ$) with the width of 20 arcsec. Bottom) For ESO~435$-$~G~016, two slices are presented, one along the angle perpendicular to the kinematic minor axis of the H{\sc i} disk (PA=$\sim112^\circ$, black line) and one through the maxima in the receding and the approaching side of the outer disk (PA=$\sim$160$^\circ$, gray line).}
% \label{fig_atcaxv}
%\end{figure}

\subsection{High resolution H{\small I} morphology and kinematics}
The ATCA H{\sc i} channel maps and the intensity maps are shown in Figures~\ref{fig_atcach} and \ref{fig_atcamo}. In the high resolution ATCA imaging data, we do not find any continuous H{\sc i} gas emission between the pair. However, resolved structures reveal detailed features associated with individual galaxies. 

Both galaxies are found with a very extended gas disk. The largest feature is more extended than $3\times D_{25}$, where $D_{25}$ is the optical size measured at 25$-$mag~arcsec$^{-2}$ in $B-$band \citep[RC3]{RC3}. This relative extent is comparable to those of BCDs located in low-density environments \citep[e.g.,][]{vanZee98}. The H{\sc i} gas is quite symmetric out to a few times of $D_{25}$, with the peak coinciding with the optical center within $<1~$kpc. In the outskirts of the gas disk however, the gas morphology becomes highly irregular and asymmetric. ESO~435$-$IG~020 shows a long gas tail of $\sim11~$kpc, pointing to the southeast. The H{\sc i} flux along the tail is measured to be $\sim20$\% of the total flux of this system. In the case of ESO~435$-$~G~016, a very extended structure is found mostly on the east side of the galaxy, with a projected size of $\sim10\times36~$kpc$^2$. The local H{\sc i} peak in this feature is $\sim3\times10^{19}~$cm$^{-2}$, well below the star formation threshold \citep{KS_law,bigiel08}. The H{\sc i} flux outside the symmetric gas disk is $\approx23$\% of the total flux measured from this galaxy. 

The total H{\sc i} mass of individual systems is calculated using the following expression,
\begin{equation}
  M_{\rm {HI}}~[M_\odot]= 2.356 \times 10^5 S_{\rm {HI}}~d^2,
\end{equation}
where $S_{\rm{HI}}$ is the integrated H{\sc i} flux in Jy~km~s$^{-1}$ and $d$ is the distance to the galaxy in~Mpc. We adopt a distance of 9.0~Mpc and 9.1~Mpc for ESO~435$-$IG~020 and ESO~435$-$~G~016, respectively \citep{11HUGS_Ha}. The total H{\sc i} masses measured from the ATCA data are $7.6\times10^7~M_\odot$ and $3.1\times10^8~M_\odot$, yielding $M_{\rm HI}/L_B$ of 0.38 and 0.52 for ESO~435$-$IG~020 and ESO~435$-$~G~016. These are comparable to $M_{\rm HI}/L_B$ found in the BCD population \citep{hucht07}, but larger than normal spirals by a few factors \citep[e.g., $M_{\rm HI}/L_B\approx0.21$ for the entire sample of the H{\sc i} Nearby Galaxy Survey (THINGS), and smaller when dwarf galaxies are excluded;][]{THINGS08}.

The gas peculiarities of the pair are also seen in their kinematics. At this velocity resolution, the H{\sc i} disk of ESO~435$-$IG~020 barely shows rotation across the morphologically symmetric part. Along the gas tail however, the velocity changes quite steeply, exceeding the velocity range found in the main H{\sc i} disk as shown in Figure~\ref{fig_atcamo}. The tail is largely responsible for the linewidth of this galaxy. 

Meanwhile, ESO~435$-$~G~016 shows clear rotation across the stellar disk, and out to a few optical radii. Within the stellar disk, the kinematical major axis measured using the tilted ring model is 112 degrees in PA, being more or less consistent with the optically defined major axis (117 degrees) but with the H{\sc i} peak offset from the optical center (see Table~\ref{tbl_hiprop}). Beyond the optical disk however, the velocity structure becomes warped, with the outer disk PA deviating from the inner disk by $\sim$50 degrees (PA$\sim$160 degrees). Across the eastern extension, the velocity changes by $>\sim50$~km~s$^{-1}$, with the same velocity gradient as the receding side as well as the tail of ESO~435$-$IG~020, as seen in Figure~\ref{fig_atcamo}. Unlike the tail of the other galaxy however, this extension does not exceed the velocity coverage of the main H{\sc i} disk. If the gas is on its way out, the stripping may have started recently and the gas has not been accelerated more than the escape velocity of the galaxy. On the other hand, if the gas is of external origin, the accretion might have been going for a while to require a similar velocity range to the main gas disk.  Alternatively, the gas which had been stripped from the past interaction might be currently falling back. 

The H{\sc i} linewidths are measured at the velocities where the H{\sc i} flux drops to 20\% and 50\% of the peak in H{\sc i} profile for each galaxy (Figure \ref{fig_hiprof}). Neither has a linewidth measured at 50\% of the peak larger than 150~km~s$^{-1}$, implying that these galaxies are unlikely to have gone through a major merging recently \citep{sung02}. However, this still does not rule out the possibility of minor merging or tidal perturbation, not only by other neighbors but also by each other.

H{\sc i} measurements including H{\sc i} mass, linewidth, and the velocity can be found in Table~\ref{tbl_hiprop}. The H{\sc i} velocity has been measured both across the entire H{\sc i} disk, and within the main disk of symmetric morphology.

\section{Discussion}

\subsection{Intergalactic gas between the pair}
In spite of a number of morphological and kinematical peculiarities in H{\sc i} associated with both galaxies, we do not find any H{\sc i} gas emission between the pair in the ATCA data down to  $\approx 10^{19}$~cm$^{-2}$ at the 3$-\sigma$ level (assuming the linewidth of 50~km~s$^{-1}$). Then the question is whether the H{\sc i} gas bridge revealed by the HIPASS survey is real, or is it simply the HIPASS resolution which makes them appear to be connected. In fact, the total ATCA flux is only $\sim$61\% of the HIPASS flux, and it is possible that large-scale structures with low density between the pair have been resolved out in the ATCA observations. If the H{\sc i} gas between the pair found in the HIPASS data is indeed real rather than the results of an overlap between two unresolved gas blobs, we might be able to extract the intergalactic H{\sc i} gas from the HIPASS data.

In order to verify the presence of H{\sc i} gas between the pair, and also to assess its flux, we have compared the HIPASS and the ATCA image in the same resolution as the HIPASS data, using the following procedure. First, the ATCA image was convolved to the same resolution as the HIPASS image, from 218 $\times$ 85 arcsec$^2$ to 930 $\times$ 930 arcsec$^2$. Then the convolved ATCA image was re-gridded to the same cell size as the HIPASS image from 20 arcsec to 240 arcsec.

\begin{figure}
 \begin{center}
 \includegraphics[bb=70 170 525 682,width=0.5\textwidth]{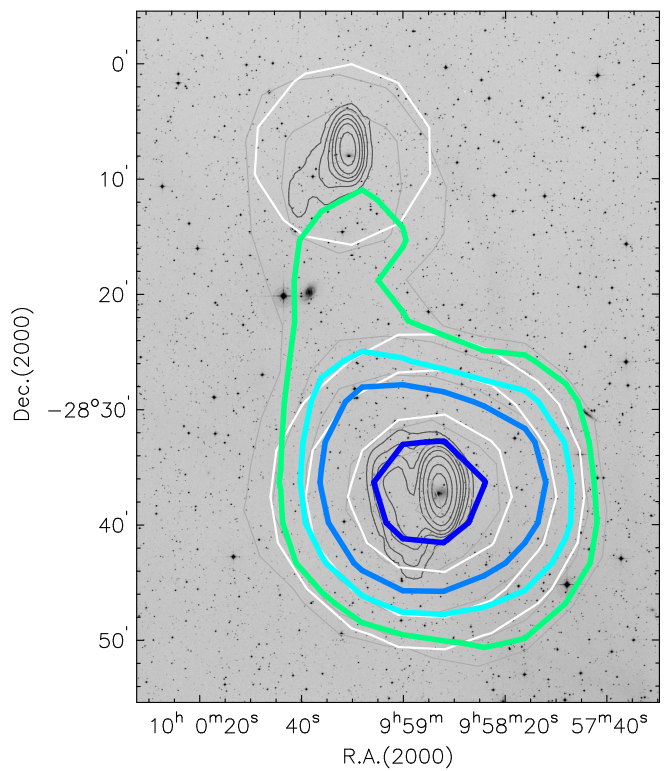}
 \end{center}
 \caption{HIPASS and ATCA H{\sc i} maps are represented by contours overlaid on the DSS red optical image with the residual emission between them. The contours of the HIPASS and the high-resolution ATCA images are shown in light- and dark-gray, respectively. The levels are the same as in Figures~\ref{fig_hipass} and \ref{fig_atcamo}. The smoothed and regridded ATCA image is shown in white contours at the same levels as the HIPASS data. The residual emissions of (1, 2, 3, 5) $\times$ $\sigma$ are shown in green, cyan, light-blue, and blue, respectively, which are comparable to (2.5, 5, 7.5, 12) $\times$ $\sigma$ of the HIPASS data.}
 \label{fig_residu}
\end{figure}

The result is presented in Figure \ref{fig_residu}. 
%The original ATCA and HIPASS data are shown in dark- and light-gray contours, respectively. The convolved and re-gridded ATCA image is shown in white contours of the same levels as the HIPASS contours. The residual emission is shown by thick lines: green, cyan, light blue, and blue represent 1, 2, 3, and 5$-\sigma$ of the residual image, which are comparable to 2.5, 5, 7.5, and 12$-\sigma$ of the HIPASS data.\LEt{ this paragraph is largely repetitive of information given in the figure heading - please cut or rephrase to present new or relevant information.} 
The H{\sc i} emission centered on ESO~435$-$IG~020 in the convolved ATCA data is offset by $\approx$ 2.4 arcmin from the nearest local peak around this galaxy in the HIPASS data. Considering that the positional accuracy of the HIPASS survey is $\sim$1 arcmin or so \citep{zwaan04}, this offset is likely to be real. However, as previously mentioned, the total flux measured from the ATCA data is consistent with the flux from the single-dish measurement, and hence the residual around this galaxy is below 1$-\sigma$ of the residual image.

Instead, most of the missing flux is found around ESO~435$-$~G~016, as expected from the comparison of previous single-dish measurements and our ATCA fluxes of individual galaxies (see Section 4.1). The total residual flux that is likely to be associated with this galaxy (i.e., Dec. $\lesssim$ $-$28$^\circ$ 24$'$) accounts for $\sim$89\% of the missing flux. Intriguingly, the remaining $\sim$10\% is found in the intergalactic space between the lowest contours around the two galaxies ($-$28$^\circ$ 24$'$ $\lesssim$ Dec. $\lesssim$ $-$28$^\circ$ 15$'$). Although the net residual emission around ESO~435$-$IG~020 is insignificant, the offset of the HIPASS peak to the south from that of the ATCA data also supports the possibility that the intergalactic H{\sc i} gas between the two galaxies is likely to be real.
 
\subsection{Star formation histories}

\begin{figure}
 \begin{center}
 \includegraphics[bb=0 275 610 570,width=0.5\textwidth]{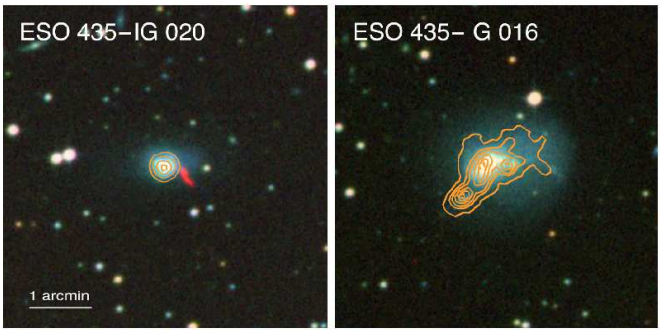}
 \end{center}
 \caption{FUV contours (orange) are overlaid on the SuperCOSMOS UKST $Bj$, $R$, and $I$ 3-color image. A scale bar of 1 arcmin is shown on the left bottom corner, which corresponds to $\approx2.6~$kpc at a 9~Mpc distance. The red tail in ESO~435$-$IG~020 is an imaging artifact in the $I-$band observations.}
 \label{fig_skymap}
\end{figure}

We study the star formation properties of this pair of BCDs in this section using archival optical, ultraviolet, and infrared observations. The far-ultraviolet (FUV) and near-ultraviolet (NUV) imaging data of our targets are obtained from the {\em Galaxy Evolution Explorer} (GALEX) UV satellite telescope's archive.  Fortunately, there exists deep GALEX observations for this pair of galaxies. The integrated $FUV-NUV$ colors (corrected for Galactic extinction) for ESO~435$-$~G~016 and ESO~435$-$IG~020 are 0.65 and 0.43 in AB magnitudes, respectively. Mid- and far-infrared measurements are obtained using archival observations from the Spitzer and Akari space telescopes. The multi-wavelength flux density measurements used in this section can be found in Table~\ref{tbl_archiv}. Figure~\ref{fig_skymap} overlays the FUV observations (orange contours) on the optical 3-color image derived from SuperCOSMOS \citep{hambly01}.

\begin{table}
\caption{Multi-wavelength observations used in SED fitting.}
\label{tbl_archiv}
\small{
\begin{center}
\begin{tabular}{lcc}
\hline
\hline
 & ESO 435$-$IG 020 & \\
Band & Flux density (Jy) or luminosity & Reference \\
\hline
FUV (1515 \AA) &   $1.91 (\pm 0.19)\times 10^{-3} $ & 1\\
NUV (2273 \AA)&    $3.22 (\pm 0.32)\times 10^{-3} $ & 1 \\
H$\alpha$ (luminosity) & $1.25~\times~10^{40}$ (ergs s$^{-1}$) & 2 \\
$B$ &  $6.99 (\pm 0.69)\times 10^{-3} $ & 3 \\
$R$ &  $1.13 (\pm 0.11)\times 10^{-2} $ & 3 \\
$J$ & $1.15 (\pm 0.05)\times 10^{-2} $ & 4 \\
$H$ & $1.07 (\pm 0.08)\times 10^{-2} $ & 4 \\
$K$& $ 6.89 (\pm 0.91)\times 10^{-3} $ & 4 \\
3.6 $\mu$m & $ 5.72 (\pm 0.17)\times 10^{-3} $  & 5 \\
4.5 $\mu$m& $ 3.91 (\pm 0.12)\times 10^{-3} $   & 5 \\
5.6 $\mu$m & $ 2.32 (\pm 0.11)\times 10^{-3} $   & 5 \\
8.0 $\mu$m & $ 5.68 (\pm 0.19)\times 10^{-3} $   & 5 \\
24 $\mu$m & $ 7.59 (\pm 0.16)\times 10^{-2} $   & 5 \\
70 $\mu$m & $ 9.35 (\pm 0.47)\times 10^{-1} $   & 5 \\
160 $\mu$m & $ 5.31 (\pm 0.76)\times 10^{-1} $  & 5 \\
\hline
 & ESO 435$-$ G 016 & \\
\hline
FUV (1515 \AA) & $1.08 (\pm 0.11)\times 10^{-3} $ & 1\\
NUV (2273 \AA)&  $2.15 (\pm 0.22)\times 10^{-3} $ & 1\\
$B$ &  $1.81 (\pm 0.16) \times 10^{-2} $   & 6 \\
$R$ &  $3.88 (\pm 0.34) \times 10^{-2} $   & 6 \\
$J$ & $6.11 (\pm 0.27) \times 10^{-2} $ & 4 \\
$H$ & $7.09 (\pm 0.35) \times 10^{-2} $ & 4 \\
$K$ & $5.70 (\pm 0.39) \times 10^{-2} $ & 4 \\
65 $\mu$m &   3.79 ($\pm 1.33$)   &  7\\
90 $\mu$m &   2.13 ($\pm 0.64$)    & 7 \\
140 $\mu$m &  4.55 ($\pm 2.73$)       &7 \\
\hline
\hline
\end{tabular}
\end{center}}
Key to Column 3: 1 -- remeasured by this work; 2 -- \citet{11HUGS_Ha}; 3 -- \citet{BCDatlas}; 4 -- \citet{skrutskie03}; 5 -- \citet{engelbracht08}; 6 --  \citet{lv89}; 7 -- \citet{pollo10}.
\end{table}

\begin{figure*}
 \begin{center}
 \includegraphics[bb=55 365 620 870,clip,width=0.49\textwidth]{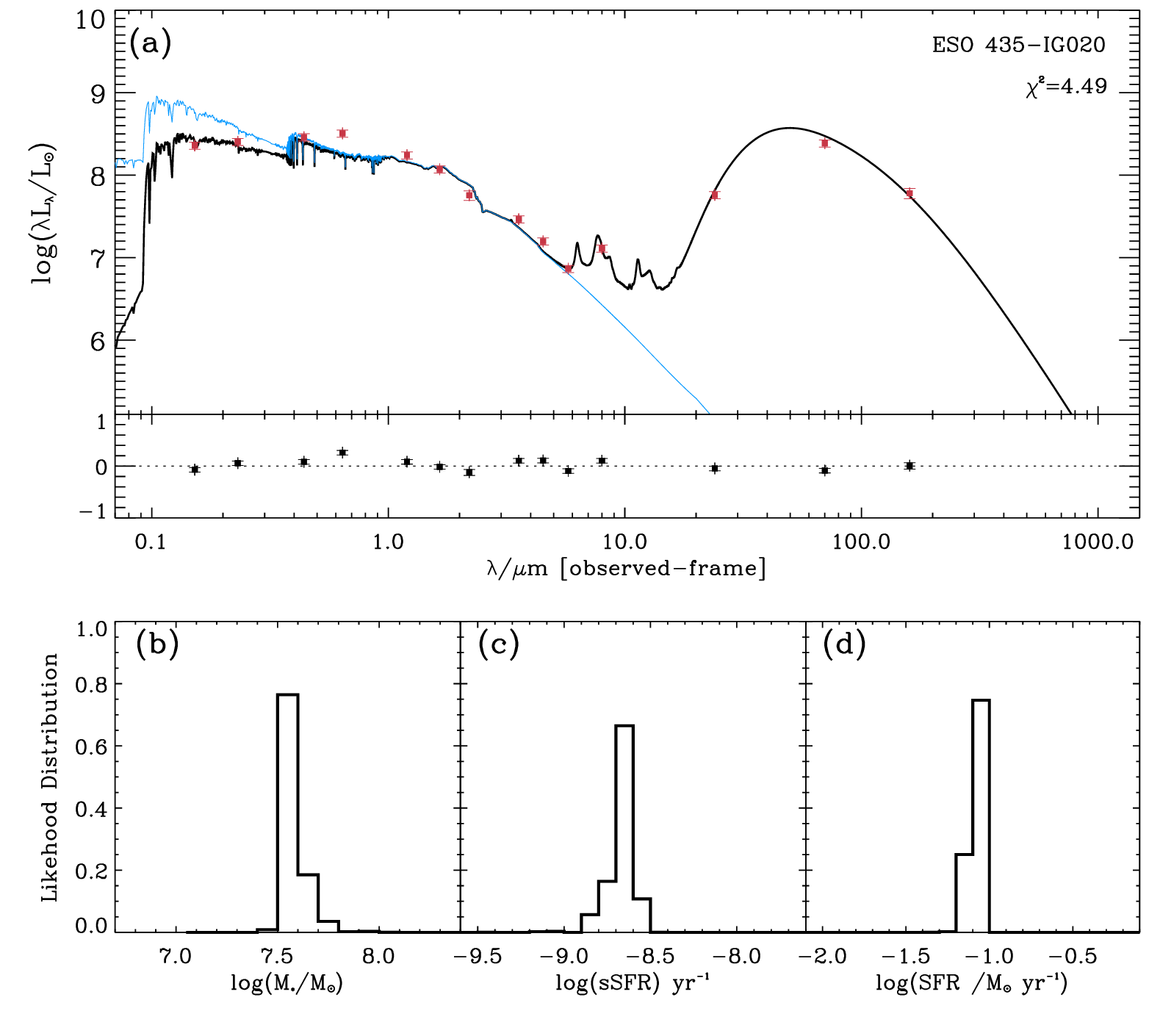}
 \includegraphics[bb=55 365 620 870,clip,width=0.49\textwidth]{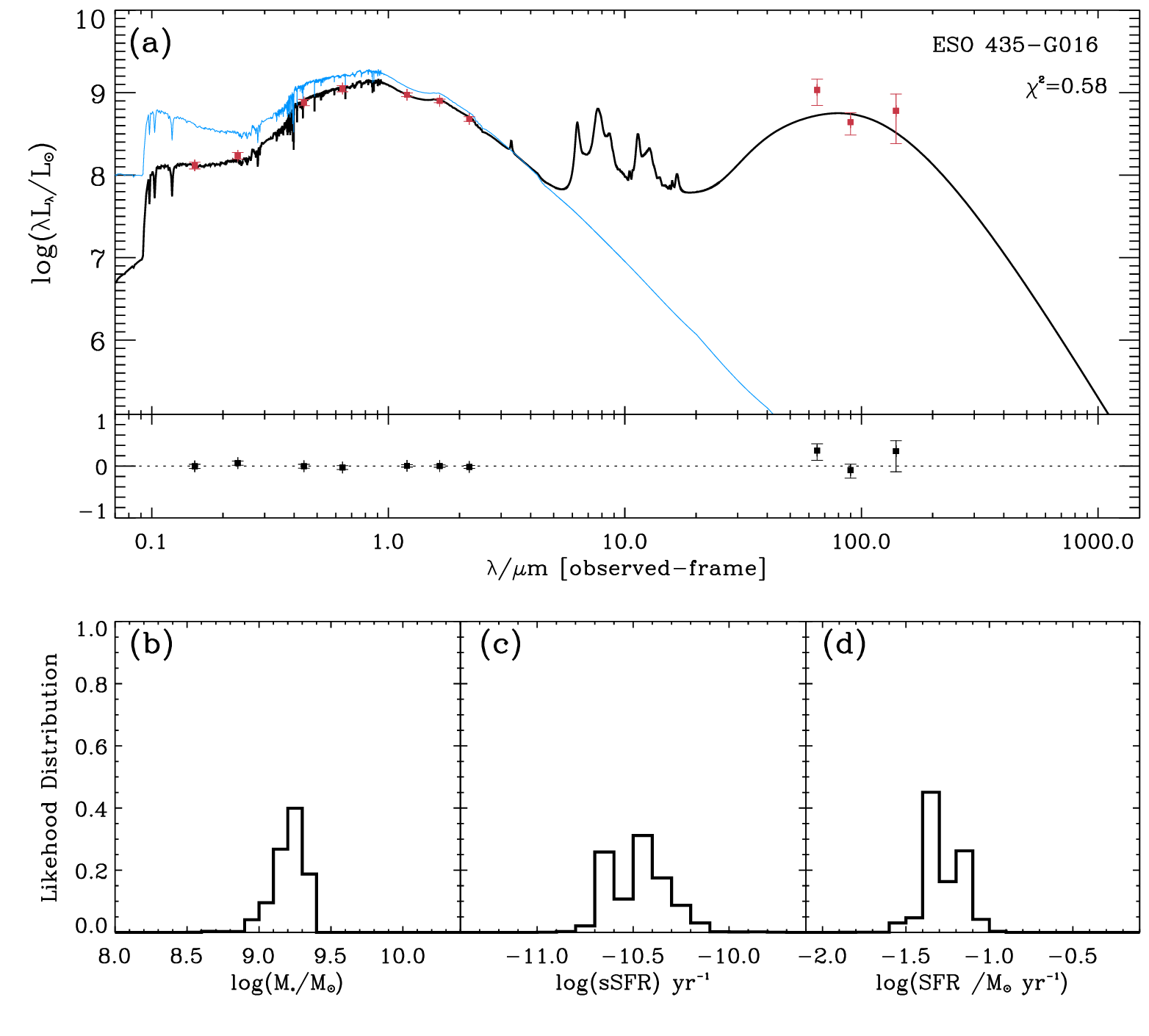}
 \end{center}
 \caption{Observed multi-wavelength SED of ESO 435$-$IG 020 and ESO 435$-$ G 016 are shown as red points in panels (a) on the left and on the right, respectively. The thick black solid lines represent the best fitting SED model estimated by MAGPHYS and the thin blue solid line represents the spectrum for the unattenuated stellar population. The goodness-of-fit is indicated by the $\chi^2$ value shown in panels (a). For both galaxies, panels (b), (c), and (d) show the resulting likelihood distributions for the stellar mass, specific star formation history (sSFR), and the star formation rate (SFR) for each of the modeled SED.}
 \label{fig_sedgal}
\end{figure*}

Due to the complex evolutionary history for this pair of BCDs, single stellar population models are insufficient for characterizing the star formation history of such galaxies.  As such, we approximate and constrain the star formation history by fitting the observed spectral energy distributions (SED) from the far-ultraviolet to the far-infrared to model star formation histories using the Multi-wavelength Analysis of Galaxy Physical Properties software \citep[{\tt{MAGPHYS}}; ][]{dacunha08}.

The observed multi-wavelength SEDs of ESO~435$-$IG~020 and ESO~435$-$~G~016 are shown as red points in panels (a) of Figure~\ref{fig_sedgal}. 
%The thick black solid lines represent the best fitting SED model estimated by {\tt{MAGPHYS}} and the thin blue solid line represents the spectrum for the unattenuated stellar population.  The goodness-of-fit is indicated by the $\chi^2$ value shown in panel (a). Panels (b), (c), and (d) show the resulting likelihood distributions for the stellar mass, specific star formation history (sSFR), and the star formation rate (SFR) for each of the modeled SED.\LEt{ This paragraph is just a repetition of the heading to Figure 7 - please cut or rephrase to include new and relevant information.} 
For ESO~435$-$IG~020, the greatest difference between the observed SED and the model fit occurs in the $R-$band.  As can be seen on the left of Figure~\ref{fig_sedgal}, the observed $R-$band flux density is greater than that from the model.  We attribute this discrepancy in the $R-$band to the strong H$\alpha$ emission that has been observed in this galaxy. Previous studies \citep[e.g., ][]{11HUGS_Ha} find the H$\alpha$ luminosity to be 1.2~$\times$~10$^{40}$~ergs~s$^{-1}$.  According to the best fitting {\tt{MAGPHYS}} model, ESO~435$-$IG~020 is consistent with a very low stellar mass dwarf galaxy that is currently forming stars with high efficiency as indicated by the sSFR value of 4.5~$\times$~10$^{-8}$~yr$^{-1}$.  For comparison, the average sSFR value for nearby star-forming galaxies is 3.2~$\times$~10$^{-9}$~yr$^{-1}$ \citep{wong16}.  Therefore, ESO~435$-$IG~020 is more efficient at forming stars by nearly an order of magnitude relative to other star-forming nearby galaxies.  The model finds that it has been approximately 1.7 Gyr since the last burst of star formation in this galaxy. Similarly, the H{\sc i}-normalized star formation efficiency ($SFE_{\rm{HI}}=SFR$/$M_{\rm{HI}}$) for ESO~435$-$IG~020  is approximately four times greater than the average found in nearby star-forming galaxies \citep{wong16}.

Conversely, the best fit models find that ESO~435$-$~G~016 is likely to be 50 times more massive in stellar mass. ESO~435$-$~G~016 has an average star formation rate over the last 10 Myr and 2 Gyr that are factors of 2.3 and 4.2 less than that for ESO~435$-$IG~020, respectively. The best fit model suggests that it has been 5.3 Gyr since the last burst of star formation ended for ESO~435$-$~G~016. The best fit model finds a median sSFR that is a factor of 2.7 less than the average nearby star-forming galaxy of similar stellar mass.  The $SFE_{\rm{HI}}$ for ESO~435$-$~G~016 is very similar to the average value found for nearby star-forming galaxies  \citep{wong16}.

Further constraints on the star formation history on shorter timescales (a few Myr) are possible via the comparison of H$\alpha$ to FUV observations. However, archival H$\alpha$ observations are not available for ESO~435$-$~G~016. As such, we can only directly confirm that there is a significant population of very young stars ($<$10 Myr) in ESO~435$-$IG~020. This is consistent with the strong ``bursty'' star formation history that we found for ESO~435$-$IG~020 via SED modeling. Future H$\alpha$ observations of ESO~435$-$~G~016 will be useful for estimating its star formation history on shorter timescales. However, we can already tell from the SED modeling and sSFR that it is less bursty and less intense than that of ESO~435$-$IG~020. Nevertheless, what makes ESO~435$-$~G~016 still interesting is its kinematical structures which indicate that this system is currently under the influence of the tidal force - potentially the origin of its recent active star formation, as described in the following section.

\subsection{Interaction history}

As argued in 5.1, the presence of intergalactic H{\sc i} gas between the pair is somewhat likely. Also, as seen in the high-resolution ATCA data, both galaxies reveal a number of peculiarities in their H{\sc i} distribution as well as in their optical morphology. All these strongly suggest that they have been tidally disturbed, potentially due to each other but also possibly by other neighbors. 

In order to investigate the feasibility of tidal interaction with other neighboring galaxies, we probe the environment around the pair of $\sim$0.5~Mpc radius within $\pm$300~km~s$^{-1}$, that is, a comparable size and velocity coverage of a typical galaxy group like our own. Within given ranges, there are 12 optically identified galaxies and several galaxy groups. 
%as shown in Figure~\ref{fig_neigal}.
Both ESO~435$-$IG~020 and ESO~435$-$~G~016 are classified as the members of the NGC~3056 triplet, which is located in the outskirts of the NGC~3175 group \citep{mk09,mk11}. Among these, we find only two sizeable galaxies that could have tidally perturbed the pair, NGC~3056 and NGC~3113. In $I-$band they are brighter than ESO~435$-$~G~016 by $\sim$0.9~mag and $\sim$1.3~mag, respectivey \citep{doyle05,springob07}.

% within the zero-velocity sphere \citep{mk11} of ESO~435$-$~G~016/020, which is $\sim0.27~$Mpc in radius, taking 10$^{10}~M_\odot$ as an upper limit of the dynamical mass of the pair (cf. the HIPASS flux and the $B-$band luminosity of the pair yields the total baryon mass of the pair of $\sim9\times10^8~M_\odot$). 

The distances from NGC~3056 and NGC~3113 to the center of the pair are $\sim154~$kpc and 196~kpc, assuming the same distance to the pair, that is, 9~Mpc. Adopting 300~km~s$^{-1}$, a typical galaxy speed in groups, the crossing time between the pair and these two relatively massive galaxies is $480-640$~Myr or so. This is shorter than the timescale for the last burst of star formation which is likely to be responsible for the dominant young stellar population in both galaxies ($>$~1 Gyr). This implies that any of these neighbors is potentially responsible for the last burst of star formation, if it was caused by tidal interaction.

However, for a crossing time between these neighbors and our BCD galaxies to be comparable to the age of the youngest stellar population of ESO~435$-$IG~020 that is traced by H$\alpha$, they should be moving with a relative velocity of an order of $\sim1000~$km~s$^{-1}$, which is unlikely in this environment. Therefore any of the close neighbors, including the two most massive galaxies, are unlikely to be the most recent active star formation in our BCD galaxies.

Instead, the interaction between the pair seems to be a more feasible scenario. Assuming the same velocity as their neighbors, the crossing time between these two BCD galaxies is $\sim$260~Myr, still too large to explain their recent active star formation. If the past encounter between the BCD pair was a fly-by however, its impact might have been just strong enough to pull out the materials from the outer disk, mostly gas, without triggering strong starbursts in the inner region of the galaxies. Since the close approach, some of the tidally stripped gas may have been recaptured \citep[e.g.,][]{dr13}. Then stars can be actively formed by the reaccreted gas directly funneling into the galaxy than by generating shocks {into the halo gas of the host galaxy}, which is more likely for low mass systems \citep[e.g.,][]{benson10}.

At the same time, this accretion-induced star formation activity can be aided by the torque and shear on the gas. Indeed, the kinematical structure of ESO~435$-$~G~016's H{\sc i} disk is quite intriguing. While the position angle (PA) of the inner H{\sc i} is more or less consistent with the optically defined position angle (PA), the angle connecting the local velocity maxima in the receding and approaching side appears to be tilted counterclockwise from the inner PA. In order to measure the offset more quantitatively, we modeled a cube of a rotating disk using {\tt GALMOD}, a package implemented in {\em GIPSY}\footnote{{Groningen Image Processing System,\\ \tt https://www.astro.rug.nl/$\tilde{~}$gipsy/}}.
%Since {\tt GALMOD} is designed to construct a model based on an axially symmetric H{\sc i} distribution, mainly the west part of ESO~435$-$~G~016 out to a few optical radii around the optical center, can be reliably modeled. 

As a result, the modeled kinematic structure based on the optical morphology in the inner region of the main H{\sc i} disk agrees well with the observed velocity field, as indicated by an ellipse and a gray line along the optical major axis in Figure~\ref{fig_atcamo}. In the outer disk however, a second rotating component is present along the axis tilted counterclockwise from the inner kinematic axis. The position angle of the second kinematic component measured from the model, PA$\approx$160$^\circ$, well matches the axis connecting the local velocity maxima in the outer H{\sc i} disk, as shown by the second gray line in Figure~\ref{fig_atcamo}.

\begin{figure}
 \begin{center}
 \includegraphics[bb=5 5 450 350, width=0.48\textwidth]{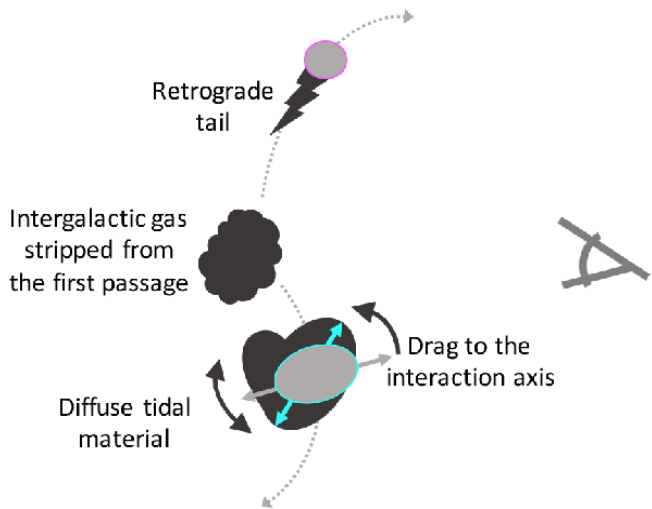}
 \end{center}
 \caption{A possible interaction scenario of the pair. A few 100 Myr ago, the two galaxies may have gone through a close encounter, tidally pulling out the gas from their outer disks. The intergalactic H{\sc i} hinted by the HIPASS data must be part of the tidally stripped gas. Based on the direction of the gas tail of ESO~435$-$IG~020 and the star formation/interaction timescale between the pair, two galaxies are inferred to be moving away from each other rather than being on the way to the second encounter. However, the main kinematic axis might still be dragged toward the interaction counterpart, generating torque on the disk (more effectively on the gas component), which can be clearly seen only in ESO~435$-$~G~016 where the structure is better resolved than ESO~435$-$IG~020, at the resolution probed in this work. The recent bursts of star formation are likely to have been triggered by the shear on the gas disk and/or the re-accretion of the stripped gas.}
 \label{fig_sketch}
\end{figure}

The offset between the inner and the outer kinematic axes might have been caused by the interaction with ESO~435$-$IG~020, which could induce the shear. The speculation that the outer gas disk is likely to have been tidally dragged counterclockwise is also supported by a recent ALMA (Atacama Large Millimeter Array) observation by \citet[][AM~0956$-$282 in their sample]{ueda14}. ESO~435$-$~G~016 is detected in $^{12}$CO ($J=1-0$) in the inner $\sim1.6~$kpc radius. Intriguingly, the CO disk also has two position angles. The outer part of the CO disk, which is composed of the northwest extent and the southeast cloud, is more tilted to the counterclockwise in comparison to the inner CO disk. Especially cool gas components are expected to be influenced more effectively by torque due to their low velocity dispersion and collisional dynamics \citep[e.g.,][]{bour11}. Hence the fact that the kinematical axis of the outer CO disk is rotated counterclockwise like the outer H{\sc i} disk, is strongly suggestive of the presence of such force in this galaxy. \citet{ueda14} also propose that the southeast CO cloud is likely to have been ejected from the main molecular gas disk due to the interaction.

In Figure~\ref{fig_sketch} we illustrate a possible interaction scenario of the pair based on all the observational evidence and the environment of the pair. In this scenario, the pair have gone through a close encounter a while ago ($\sim$ a few 100 Myr ago assuming typical group environment), which tidally stripped gas from the outer region. The intergalactic H{\sc i} hinted at by the HIPASS data must be part of the stripped gas from the first encounter. 

It is difficult to judge whether the orbit is closed or not, but if ESO~435$-$IG~020's tail was formed via fly-by interaction between the pair, its direction suggests that the two galaxies are in situ moving apart based on some simulations where tidal materials are found to form in retrograde after the first encounter in fly-by interactions \citep[e.g.,][]{pkdb11}. Likewise it is hard to tell whether the tail of ESO~435$-$IG~020 and ESO~435$-$~G~016's extension are being stripped or reaccreted. However, accretion is more feasible if the two galaxies are currently moving away, and their recent bursts of star formation can be also more naturally explained. 

Even if the two galaxies are moving apart as suggested by ESO~435$-$IG~020's tail, the velocity shear across the main H{\sc i} disk implies that the two galaxies are still under the influence of each other. The CO disk also shows at least two kinematic axes, with the outer position angle rotated in the same direction as the outer H{\sc i} kinematic axis, supporting the presence of torque in the gas disk. Although the shear is less visible in ESO~435$-$IG~020 due to its compact structure {and its nature that is less rotationally supported compared to ESO~435$-$~G~016,} it must be also experiencing the torque as the interaction counterpart of ESO~435$-$~G~016. This shear could be responsible for recent enhanced star formation activity in both galaxies.

Lastly, it is worth mentioning that we cannot completely rule out the merging between gas-rich dwarf galaxies as the origin of the recently enhanced star formation in these BCDs. A nucleated core and diffuse structures do well agree with BCDs originating from the merging of gas-rich galaxies simulated by \citet{bekki08}. In \citet{bekki08}'s simulation, however, only the old stars with an age $>$1~Gyr and with gas form diffuse low-surface brightness structures, and new stars mostly form in the central region after merging. However, this is not the case of our pair as shown by the FUV image in Figure \ref{fig_skymap}, especially of ESO~435$-$~G~016, where the internal structures can be better seen.

%Assuming, two galaxies have been moving on the plane of the sky for the last few tens Myrs, the velocity with respect to each other is estimated to range from $\sim$1500~km~s$^{-1}$ to $\sim$2500~km~s$^{-1}$ to arrive where they currently are. If they had the closest approach in the past and they have been moving away from each other since, the relative velocity between the pair is too high to bring them back together. Therefore it is less likely that the two galaxies have another close approach, and the past interaction could be the only fly-by event between the pair.

\section{Summary and Conclusion}
We analyzed the H{\sc i} imaging data of a BCD pair, ESO~435$-$IG~020 and ESO~435$-$~G~016, obtained using the ATCA. The goal was to find the evidence for fly-by interaction between the pair as suggested by a large H{\sc i} envelope covering both that is found by the HIPASS, and to investigate the role of tidal interaction in triggering recent active star formation in the pair. 

From high resolution H{\sc i} imaging data, we find clear indications that both galaxies have been tidally disturbed. However, intergalactic H{\sc i} gas is not detected with the sensitivity limit of the ATCA observaions, which has recovered only 61\% of the HIPASS H{\sc i} flux. Considering typical galaxy speeds in groups, there are only two candidates which are massive enough to gravitationally perturb the pair in their neighborhood. The crossing time between the BCD pair and the neighbors within a radius of a comparable size to typical galaxy groups is short enough to explain the mean age of the dominant young stellar population ($>$ 1 Gyr), while it is still too large to be responsible for the most recent active star formation event in our BCDs.

Intriguingly, the residual emission between the HIPASS and the ATCA H{\sc i} data is mainly found between the two galaxies. In addition, we found kinematical evidence for the shear on the gas disk. The second kinematic axis underlying the main H{\sc i} disk of ESO~435$-$~G~016 indicates that the gas component is under the influence of torque, which is also supported by the kinematics of its $^{12}$CO ($J=1-0$) gas disk. 

Based on our analysis, we conclude that the interaction between the two galaxies is responsible for their H{\sc i} and optical peculiarities as well as the recent star formation activities. During the most recent encounter, the pair could have lost H{\sc i} gas from the outer disk. Enhancements in recent star formation on timescales of $\sim$10~Myr might have been triggered by the reaccretion of the gas that had been stripped during the past encounter and/or the shear on the disk caused by each other.

In this study of a BCD pair, we have provided observational evidence that star formation can be enhanced by fly-by interaction between galaxies without merging. If this well-separated pair of BCDs are representative of BCDs, we propose that fly-by interactions could be one of the most important origins of BCDs. Our suggestion can explain the observation that a high fraction of BCDs are not in the merging process \citep[$\sim$60\%, e.g.,][]{sung02}.

\bigskip

\begin{acknowledgements}
We are grateful to the anonymous referee for his/her valuable comments and useful suggestions. Support for this work was provided by the National Research Foundation of Korea to the Center for Galaxy Evolution Research (No. 2010-0027910) and Science Fellowship of POSCO TJ Park Foundation. This work has been also supported by NRF grant No.~2015R1D1A1A01060516. Parts of this research were conducted by the Australian Research Council Centre of Excellence for All-sky Astrophysics (CAASTRO), through project number CE110001020. We are grateful to B${\rm \ddot a}$rbel Koribalski for her useful suggestions and comments. The Australia Telescope Compact Array is part of the Australia Telescope National Facility which is funded by the Australian Government for operation as a National Facility managed by CSIRO. This work has made use of the NASA/IPAC Extragalactic Database (NED) which is operated by the Jet Propulsion Laboratory, California Institute of Technology, under contract with the National Aeronautics and Space Administration. We thank Elisabete da Cunha and Simon Driver for their helpful guidance with {\tt MAGPHYS}.
\end{acknowledgements}


\end{document}